\documentclass[MSNbibl,nameyear,dvips]{arxstspdf}
\usepackage{flushend}
\usepackage{stfloats}
\volume{26}
\issue{1}
\pubyear{2011}
\firstpage{15}
\lastpage{16}
\doi{10.1214/11-STS337D}
\referstodoi{10.1214/10-STS337}

\begin{document}
\begin{frontmatter}

\title{Discussion of ``Statistical Inference: The~Big Picture'' by R. E. Kass}
\runtitle{Discussion}
\pdftitle{Discussion of Statistical Inference: The~Big Picture by R. E. Kass}

\begin{aug}
\author{\fnms{Robert} \snm{McCulloch}\corref{}\ead[label=e1]{robert.mcculloch1@gmail.com}}

\runauthor{R. McCulloch}

\affiliation{University of Texas at Austin}

\address{Robert McCulloch is Professor, IROM Department, University of Texas at Austin, CBA 5.202,
1 University Station, B6500, Austin, Texas 78712-0212, USA \printead{e1}.}

\end{aug}



\end{frontmatter}

Kass states (page 5) that Figure 3 is not a good general description
of statistical inference and that Figure 1 is more accurate.
I completely agree.
Kass states (page 5) that
``It is important for students in introductory courses to see the subject as a coherent, principled, whole.''
I completely agree.
Since Figure 3 represents the framework  within which statistics is
usually taught, at all levels, we have a serious problem.
These issues have bothered me deeply for a long time.
This important paper forcibly brings these  matters to light and
I~hope it is influential.

Figure 3 represents ``sampling from a finite population.''
There are a large number of unknown numbers and we randomly pick some of them
to uncover.  The ``population quantities'' are summaries of all of the
numbers.  These are the parameters.  Sample quantities are summaries of the uncovered
numbers.  All ``randomness'' arises from that which we inject by randomly picking the sample.
This is so obviously not a description of what we usually do in statistical modeling
that I am just amazed at its persistence.
``Randomness'' comes from my personal need to make a decision in an uncertain context
(Lindley, \citeyear{Lin85}).

As Kass does, let me take a ``simple'' example.
I~just taught an introductory statistics class.
How did the need for ``probabilistic thinking'' come  into
the course in a way that the students could immediately see the
need for it?
The practical problem of choosing an investment portfolio for one
period was considered.
You could put your money in a~riskless asset (government bonds) with known return
or a risky asset (the market).
Past data on market returns are available.
Several issues need to be discussed.
There is not really anything such as a~``riskless asset'' but compared
to stocks, goverments bonds are riskless.
There is a useful, but imperfect match between our model and the real world.
How much do the past returns guide us representing our uncertainty about
the unknown future return?
Suppose you hold the portfolio for two periods.
Is the return in the second period related to (independent of) the return in
the first?
How could returns be approximately independent?
Perhaps the theory of efficient markets sheds light on this.
That is how I introduced probability and (\textit{I think}) I got away with it.
I used a normal distribution to describe my uncertainty about the next return.
I graphically showed that my choice of mean and variance was
``somewhat consistent'' with the past but emphasized that I did not
have to match the past and we discussed deviations that one might want to
consider.
I would say that my approach was largely consistent with Kass' Figure 1.
I would also say that the idea of a~random sample from a finite (or infinite)
population was nowhere to be seen.

Figures 1 (and 4) are not easy ideas.
But not addressing them directly just makes it more confusing in the long run.
What you really need in a good course are relatively simple examples
which are entirely in the spirit of Figures 1 and 4.
I did use coins and dice in discussing the intuitive idea of
independence.  However, I~also emphasize that saying returns
are independent is a big assumption with serious consequences.

In more complex statistical modeling where we consider many variables,
many basic statistical issues must be considered.
Ones of general importance that come to mind
are the bias--variance trade-off in prediction and the
difference between prediction given passive observations of a system
versus predictions about the effect of an intervention in a system
that has not been done before (correlation versus causation).
These kinds of issues fit naturally into Figures 1 and 4.
I assume that the term ``conclusions'' is meant to include
predictions.

Of course, the other issue that statistical science stresses
is the quantification of our uncertainty about our conclusions.
Here, I still believe the Bayesian approach has real advantages.
Kass knows all the arguments better than I do
so I will not go through the list but I cannot resist rattling off a few.
Even teaching an introductory course, the severe deficiencies
of classically inspired approaches become apparent.
I could not see a huge difference between the frequentist and pragmatic
attitudes in Section 3.  We all know that most people treat confidence intervals
as probability intervals, which is not too bad since they often are reasonable
approximations.  In the testing area, it is well documented that serious power
calculations are scant.  If you just report a $p$-value, have you learned anything?
It is just a draw from the uniform under the null.
To compute the power you need to pick places under the alternative
which requires---a prior!

But, perhaps a more fundamental argument for the Bayesian perspective
is that it helps us think the way Figures 1 and 4 tell us to.
I know you can understand causality and bias--variance trade-off
and maybe all the key modeling ideas with frequentist reasoning
but I still feel that thinking about probability subjectively makes
it easier to think about modeling.
In this regard, some of my all-time favorites are Chapters 1 and 2
of Cowell et al. (\citeyear{Cowetal07}).
It is just a lot easier for me to think about next period's
return as my beliefs or ``some reasonable person's'' beliefs
than some repeated imaginary sequence contingent upon some
unknown ``true'' state of nature.
I~think about independence by
asking my\-self if one thing could help me guess another.
I~think about conditional independence by asking myself if I~knew
something else, then would the one thing help me guess\vadjust{\goodbreak} the other.
I~make up things that do not exist (parameters) in order to build a
story about how things are related.
I~just think this is a more \textit{realistic} way to
think along the lines of Figures 1 and 4.
I~do not know what the term ``random sample''
in the ``pragmatic interpretation'' on page 4 means in most
of the examples I~work.

Maybe I have been trying to think like a Bayesian for
a number of years and have found a~few examples where I think it works.
I have certainly also found a few examples where I have tried to construct
a useful prior in a high-dimensional space and failed.
I think I do use the subjective interpretation of probability in practice.
I say, ``Maybe not me if I had lots of time to think about it, but
a reasonable person could believe all this and if I compute a conditional
for him then he would believe that.''  Of course if the probability intervals
for things we care about consistently fail to ``cover'' we may want to
rethink all important aspects of our model.

Death to Figure 3.


\end{document}